\documentclass[sigconf]{acmart}
\makeatletter
\def\@ACM@checkaffil{
    \if@ACM@instpresent\else
    \ClassWarningNoLine{\@classname}{No institution present for an affiliation}%
    \fi
    \if@ACM@citypresent\else
    \ClassWarningNoLine{\@classname}{No city present for an affiliation}%
    \fi
    \if@ACM@countrypresent\else
        \ClassWarningNoLine{\@classname}{No country present for an affiliation}%
    \fi
}
\makeatother

\pdfoutput=1
\usepackage{balance}
\usepackage{subfigure}
\usepackage{color}
\usepackage{graphicx}
\usepackage{xspace}
\usepackage{multirow}
\usepackage{makecell} 
\usepackage{enumitem} 
\usepackage[ruled,vlined]{algorithm2e}
\usepackage{float}
\usepackage{ulem}
\newcommand{\parhead}[1]{\smallskip \noindent \textbf{#1}\hskip .1in}
\newcommand{\Parhead}[1]{\noindent \textbf{#1}\hskip .1in}

\newenvironment{strategy}[1][htb]
  {
   \begin{algorithm}[#1]%
  }{
   \end{algorithm}
   }

\ifdefined\isFinalized
\newcommand{\NewCommentType}[3]{}
\else
\newcommand{\NewCommentType}[3]{\expandafter\newcommand\csname #1\endcsname[1]{{\color{#2}{#3: ##1}} }}
\fi

\copyrightyear{2023}
\acmYear{2023}
\setcopyright{acmlicensed}\acmConference[WWW '23]{Proceedings of the ACM Web Conference 2023}{May 1--5, 2023}{Austin, TX, USA}
\acmBooktitle{Proceedings of the ACM Web Conference 2023 (WWW '23), May 1--5, 2023, Austin, TX, USA}
\acmPrice{15.00}
\acmDOI{10.1145/3543507.3583189}
\acmISBN{978-1-4503-9416-1/23/04}

\renewcommand\footnotetextcopyrightpermission[1]{}



\usepackage[override]{cmtt} 

\begin{document}


\newcommand{\sysname}{\textbf{TMC}\xspace}
\newcommand{\Geneva}{$\mathsf{Geneva}$\xspace}

\newcommand{\sectionref}[1]{$\S$\ref{#1}}
\providecommand{\myparab}[1]{\parhead{#1} }

\NewCommentType{todo}{red}{TODO}
\NewCommentType{Sadia}{green}{SN}
\NewCommentType{Van}{pink}{VT}
\NewCommentType{Chase}{orange}{CJ}
\NewCommentType{Kevin}{purple}{KB}
\NewCommentType{Phong}{blue}{NP}
\NewCommentType{dml}{violet}{dml}
\NewCommentType{nick}{cyan}{NF}


\title{Measuring and Evading Turkmenistan's Internet Censorship}
\subtitle{A Case Study in Large-Scale Measurements of a Low-Penetration Country}




\author{Sadia Nourin}
\affiliation{%
  \institution{University of Maryland}
}
\email{snourin@umd.edu}

\author{Van Tran}
\affiliation{%
  \institution{University of Chicago}
}
\email{tranv@uchicago.edu}

\author{Xi Jiang}
\affiliation{%
  \institution{University of Chicago}
}
\email{xijiang9@uchicago.edu}

\author{Kevin Bock}
\affiliation{%
  \institution{University of Maryland}
}
\email{kbock@cs.umd.edu}

\author{Nick Feamster}
\affiliation{%
  \institution{University of Chicago}
}
\email{feamster@uchicago.edu}

\author{Nguyen Phong Hoang}
\affiliation{%
  \institution{University of Chicago}
}
\email{nguyenphong@uchicago.edu}

\author{Dave Levin}
\affiliation{%
  \institution{University of Maryland}
}
\email{dml@cs.umd.edu}

\renewcommand{\shortauthors}{S. Nourin et al.}
\begin{sloppypar}

\begin{abstract}
Since 2006, Turkmenistan has been listed as one of the few Internet
enemies by Reporters without Borders due to its extensively censored
Internet and strictly regulated information control policies. Existing
reports of filtering in Turkmenistan rely on a small number of vantage
points or test a small number of websites. Yet, the country's poor
Internet adoption rates and small population can make more
comprehensive measurement challenging. With a population of only six
million people and an Internet penetration rate of only 38\%, it is
challenging to either recruit in-country volunteers or obtain vantage
points to conduct remote network measurements at scale.

We present the largest measurement study to date of Turkmenistan's Web
censorship. To do so, we developed \sysname, which tests the blocking
status of millions of domains across the three foundational protocols
of the Web (DNS, HTTP, and HTTPS). Importantly, \sysname \emph{does
not require access to vantage points in the country}. We apply
\sysname to 15.5M domains, our results reveal that Turkmenistan
censors more than 122K domains, using different blocklists for each
protocol. We also reverse-engineer these censored domains, identifying
6K over-blocking rules causing incidental filtering of more than 5.4M
domains. Finally, we use \Geneva, an open-source censorship evasion
tool, to discover five new censorship evasion strategies that can
defeat Turkmenistan's censorship at both transport and application
layers. We will publicly release both the data collected by \sysname\
and the code for censorship evasion.

\end{abstract}

\maketitle


\section{Introduction}
\label{sec:introduction}

Internet censorship by powerful nation-states threatens free and open
communication for those living within their
borders~\cite{RU-UA-conflict, Mahsa-Amini}. For decades, researchers
and practitioners have focused considerable efforts towards measuring,
understanding, and circumventing censorship across the globe, with
particular focus on the largest and most powerful censoring regimes,
like China~\cite{Barme1997, GFWatch, Tripletcensors},
Iran~\cite{Aryan:2013, anderson2013ir.throttling, IranTelegram,
Bock2020DetectingAE}, and India~\cite{india_censorship}. The methods
developed to meet the massive scales of these efforts range from
recruiting participants to deploying ``probes'' within the censoring
regimes~\cite{filasto2012ooni} to finding active ``echo
servers''~\cite{Raman2020CensoredPlanet},
VPNs~\cite{Hoang2019:I2PCensorMeasure, ICLab:SP20,
Hoang2022MeasuringTA}, or other responsive devices~\cite{quack} to
receive censored traffic.

Although previous efforts have been effective at measuring censorship
in different regions of the world, they face many challenges when it
comes to \emph{small countries}, especially those with a low Internet
penetration rate. For instance, it can be difficult or risky to
recruit local volunteers to test ``potentially censored'' websites in
repressive countries with a small population as their network probes
will likely stick out from other ``allowed'' network traffic. For
countries with a low Internet penetration rate, it is similarly
challenging to acquire in-country vantage points or identify viable
VPNs or responsive servers for remote network measurements.

In this paper, we introduce techniques for measuring and evading
censorship of countries with low Internet penetration rates, without
relying on traditional in-country resources like vantage points or
volunteers. In contrast to previous measurement techniques that
require servers or participants within a censoring nation-state, our
techniques exploit the fact that some smaller countries' censorship
infrastructure can be tricked into believing that an external host has
connected to an internal IP address, even if that IP address is not
actually in use. Bock et al.~\cite{Bock2021WeaponizingMF} used this
characteristic to launch amplification attacks; we use it to develop
techniques to measure and evade censorship.

We focus our study on Turkmenistan for several reasons. Most
importantly, Turkmenistan's Internet censorship behavior presents a
rare opportunity for scalable remote measurements to investigate
network filtering across all three foundational protocols of the Web:
DNS, HTTP, and HTTPS. Second, there have been recent reports of more
restrictive Internet policies in the country~\cite{TM-blocks-VPNs,
TM-slow-Internet}, resulting in sudden increases in the number of
clients seeking to use anonymous network relays such as Tor and I2P
since 2021~\cite{tor_metrics, i2p_metrics}. Finally, numerous
anecdotes have reported instances of some popular websites being
censored~\cite{TM-github-anecdotes, TM-blocks-DoH} while there has yet
to be a large-scale and systematic study on the country.

Motivated by these developments, our paper seeks to systematically
answer the following questions about Turkmenistan's censorship:
(1)~What websites are censored, and over what protocols? (2)~How does
the censorship infrastructure work? and (3)~How can we evade
Turkmenistan's censorship?

To answer these questions, we present the design and implementation of
\sysname, a large-scale measurement system capable of testing millions
of domains from outside a censoring nation-state without having access
to internal vantage points~(\sectionref{sec:method}). \sysname takes
advantage of an important characteristic of Turkmenistan's censorship
that is common (though not pervasive) among nation-state censors: it
employs ``bi-directional'' censorship. Bi-directional censors act on
traffic the same way regardless of whether the client or server is
within their borders. It applies to all traffic even if the connection
did not originate within the censored country. In contrast,
``uni-directional'' censors apply filtering policy solely on network
traffic originated from within their jurisdictions.

Turkmenistan's bi-directional censorship was discovered through
anecdotal accounts by many users \cite{TM-github-anecdotes}. Due to
its bi-directional nature, we can originate all the measurement
traffic from machines we control outside the country. However,
bi-directional censorship alone was not enough for us to perform our
measurements to \emph{every} IP address within Turkmenistan's borders.
To do so, we needed to develop additional, novel techniques to trick
the censor into believing we are communicating with an arbitrary IP
address---even if that address is not responsive to us---and then
detecting censorship had taken place. We summarize our empirical
findings as follows:

\begin{itemize}

\item Using \sysname, we examine the blocking status of more than
15.5M fully qualified domain names (FQDNs) and detect a total of 122K
censored domains~(\sectionref{sec:censored-domains}).

\item Using these censored domains, we reverse engineer the actual
blocklists used by Turkmenistan's filters, finding 6K over-blocking
regular expressions that can cause large collateral damage to more
than 5.4M domains unrelated to the domains that we believe
Turkmenistan intended to block~(\sectionref{sec:censored-domains}).

\item We use \Geneva~\cite{Bock2019GenevaEC}, an automated evasion
tool, to discover novel censorship evasion strategies. In addition to
finding that some evasion techniques that work in
China~\cite{Bock2021EvenCH} and Iran~\cite{Bock2020DetectingAE} also
work in Turkmenistan, we discover five new strategies that can defeat
Turkmenistan censorship at both transport and application
layers~(\sectionref{sec:circumvention}).

\end{itemize}
These contributions not only close the gap in the community's
understanding of Web censorship in Turkmenistan but also come up with
effective censorship evasion strategies that will hopefully assist in
the development of circumvention tools to bypass the country's
censorship at different layers of the network stack. The datasets
collected by \sysname and the code for evasion strategies that we
discovered will be made publicly available.
\begin{figure*}[t]
	\centering
    \subfigure[DNS filtering over UDP.]{\label{fig:dns_filter}\includegraphics[height=0.22\textwidth]{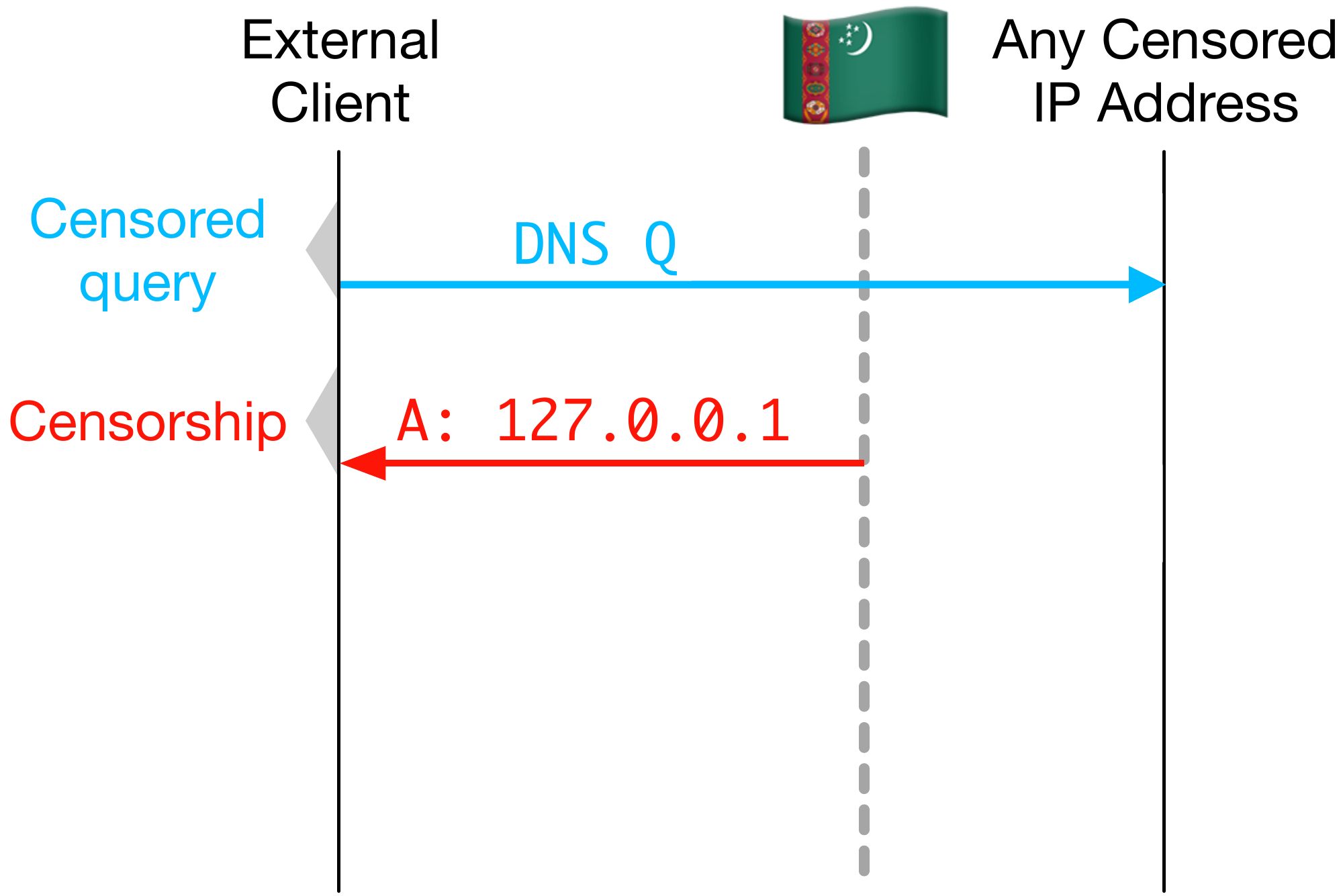}}
	\hfill
	\subfigure[Standard TCP censorship (HTTP and HTTPS).]{\label{fig:full-handshake}\includegraphics[height=0.24\textwidth]{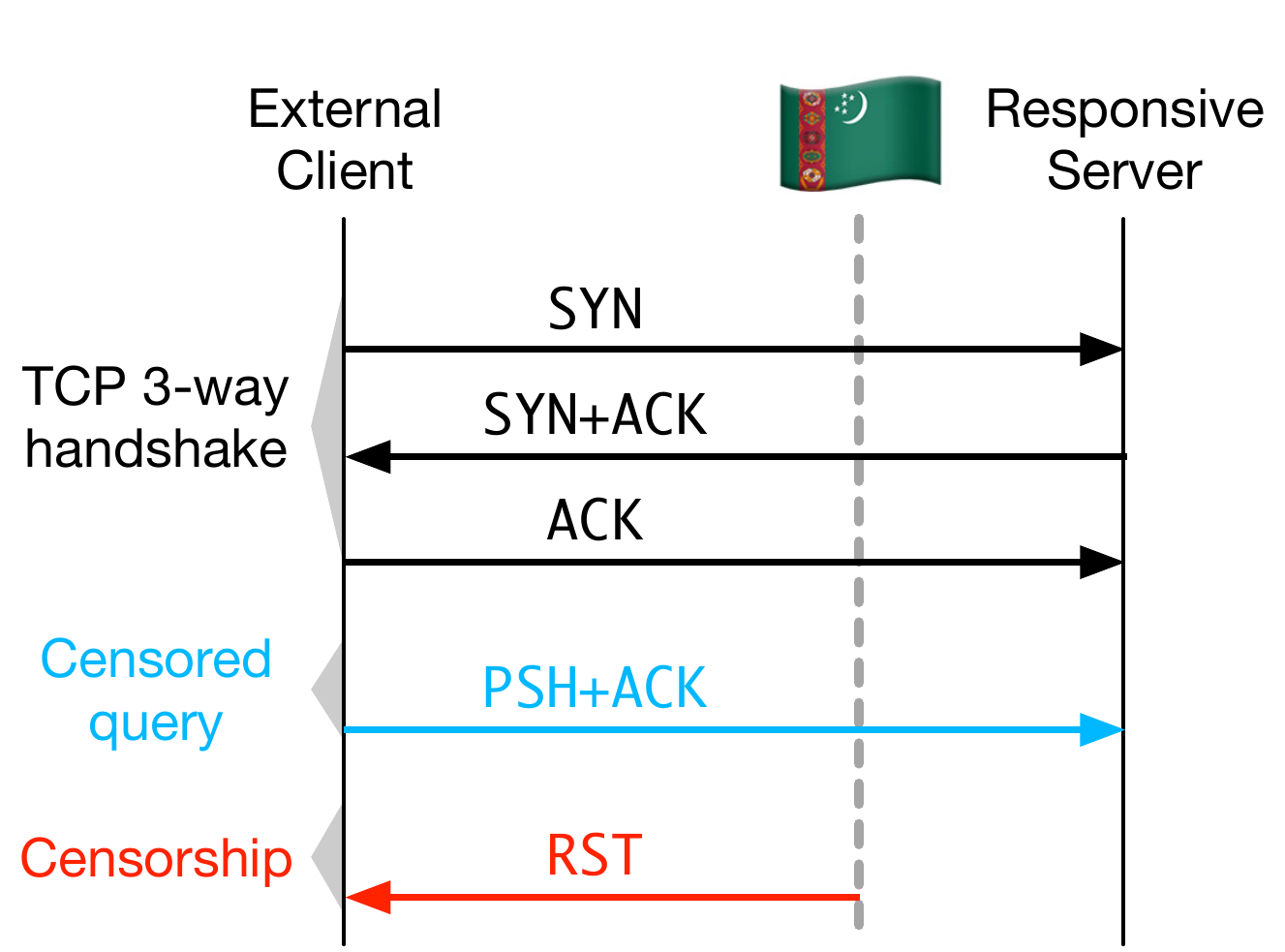}}
	\hfill
	\subfigure[Triggering TCP censorship without a server.]{\label{fig:incomplete-handshake}\includegraphics[height=0.24\textwidth]{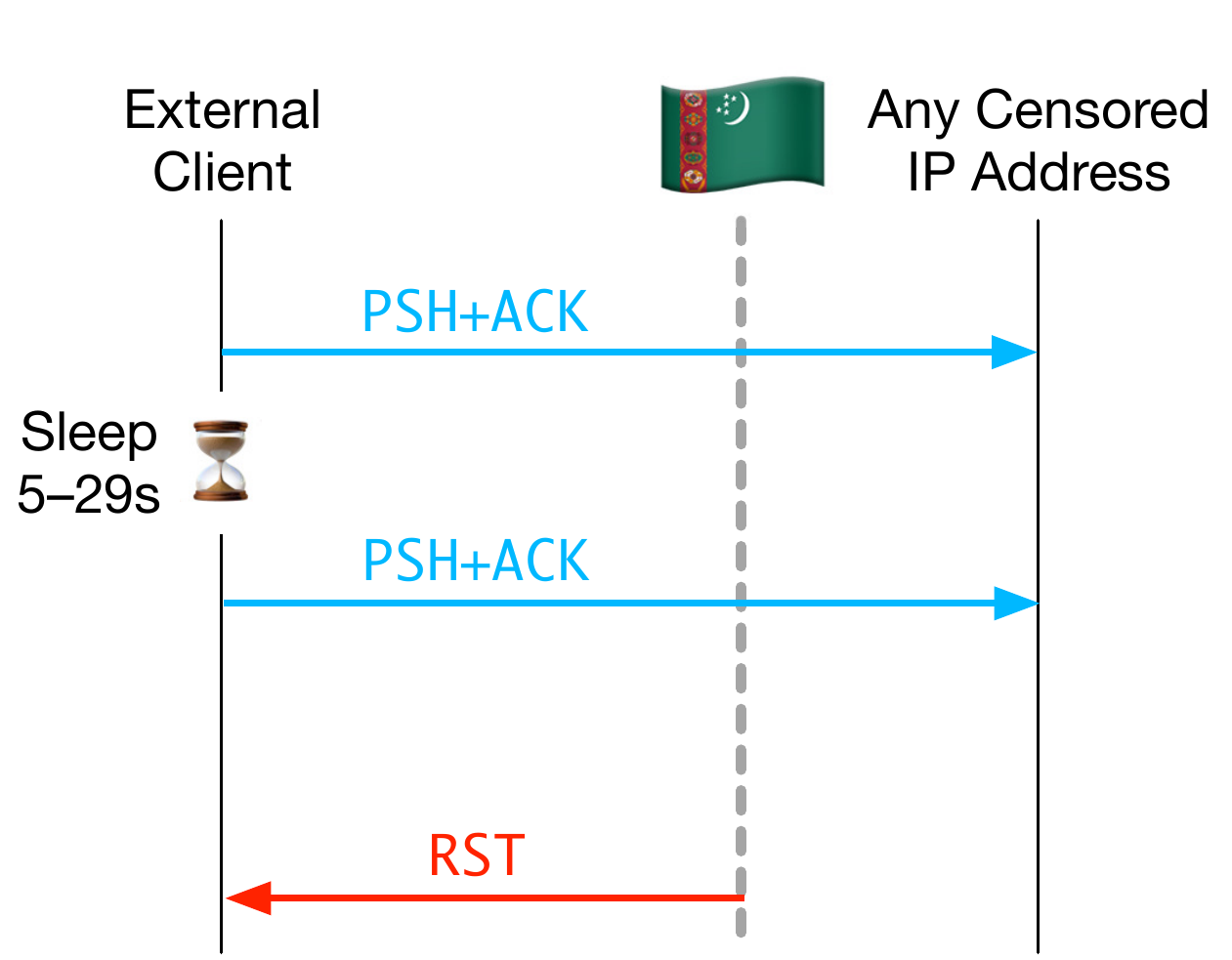}}
	\caption{Turkmenistan censorship can be triggered from outside due
	to the bidirectional blocking behavior of filtering firewalls. The
	\texttt{PSH+ACK} contains the censored domain in the GET request
	for HTTP and the SNI field for HTTPS. \sysname exploits the fact
	that Turkmenistan's firewalls can be triggered and sends a second
	\texttt{PSH+ACK} after waiting for 5-29 seconds after the first
	one that contains the censored domain.}
	\label{figure:tm_censorship}
\end{figure*}

\section{Background and Motivation}
\label{sec:background}

In this section, we first provide an overview of Turkmenistan's
information control policies. We then discuss how the country uses
different techniques for Web filtering, the challenges we initially
faced when attempting to measure censorship, and how they have
motivated us to conduct this study.

\subsection{Turkmenistan's Information Controls}
\label{sec:Turkmenistan}

From a sociopolitical perspective, Turkmenistan has a freedom score of
only 2/100 (1 is the lowest) ranked by the Freedom House in
2022~\cite{FreedomHouse2022}. This score is reflective of a series of
suppressive activities by the Turkmen government, including the
suppression of press freedom, strict control of all broadcast and
print media, as well as state-owned Internet service
providers~\cite{TM-freedom-score}.

The Turkmen government has been using different tactics to keep an
inclusive and freely accessible Internet at check. Specifically,
Internet access is astronomically expensive due to a state monopoly
while broadband speed is among the World's slowest~\cite{ONI_TM,
TM-slow-Internet}. Moreover, authorities strictly monitor all
communication channels~\cite{ONI_TM} and ban ``uncertified''
encryption software. For instance, VPN users may face a penalty of
seven years in prison~\cite{hrw_TM} and Turkmen citizens have reported
that they were required to swear on the Quran not to install a
VPN~\cite{TM-blocks-VPNs}.

\subsection{Web Censorship Mechanisms in Turkmenistan}
\label{sec:Turkmenistan-censorship}

Together with restrictive Internet regulations, the Turkmen government
also makes use of different network interference techniques for Web
censorship.
In August 2021, researchers reported that Turkmenistan was employing
significant censorship of the Net4People
community~\cite{TM-github-anecdotes}, targeting all three foundational
protocols of the Web: DNS, HTTP, and HTTPS.
Because Turkmenistan was applying this censorship in a bi-directional
manner---that is, it was censoring traffic regardless of whether it
was the client or the server inside their borders---we were able to
reproduce and understand how they were censoring each of these three
protocols:

\parhead{DNS}
DNS tampering works by taking advantage of the race condition of the
DNS protocol~\cite{rfc1034} (when a query is sent over UDP) to inject
a forged DNS response containing wrong resource records of the domain
being queried. To trigger a DNS injection from outside Turkmenistan,
one only needs to send a DNS query containing a DNS Query for a
censored domain (e.g., \texttt{twitter.com}) to an IP address located
inside the country. The censor will inject a DNS response packet
containing \texttt{127.0.0.1} as the resource record for the censored
domain (see Figure~\ref{fig:dns_filter}).

\parhead{HTTP}
For HTTP blocking, we can test if a domain is censored by initiating a
TCP connection with an HTTP server followed by an \texttt{HTTP GET}
request containing a censored domain in the \texttt{Host} field. Upon
detecting the censored domain, the censor will inject one \texttt{RST}
(reset) packet to tear down the connection (see
Figure~\ref{fig:full-handshake}).

\parhead{HTTPS}
HTTPS censorship can be triggered in much the same way. First, we can
complete a TCP connection with an HTTPS server located inside
Turkmenistan. Then, in the very next \texttt{PSH+ACK}
packet---corresponding to the TLS Client Hello message---we set the
Server Name Indication (SNI) field to a censored domain.
This causes the Turkmen censor to inject a \texttt{RST}, also shown in
Figure~\ref{fig:full-handshake}.

\medskip \noindent
Unlike the original Net4People report, we find that censorship for all
three protocols is not restricted to the protocols' traditional ports.
For example, although HTTP traditionally runs on port 80, we can
trigger HTTP censorship to \emph{any} destination port.

We find that both HTTP and HTTPS filters exhibit \emph{residual
censorship}~\cite{Bock2021Residual}.
After triggering censorship by including a censored domain in the
HTTP \texttt{Host} header or in the TLS SNI field, \emph{any}
subsequent packet matching the same TCP four-tuple (source IP:port,
destination IP:port) will cause the censor to inject a \texttt{RST}
packet.
We determined that this residual blocking behavior stops after 30
seconds from the last injected packet.

\subsection{Measurement Challenges}
\label{sec:Challenges}

Despite the ease confirming Turkmenistan's bidirectional censorship,
we face several challenges when trying to examine censored domains
\emph{at scale}.

\myparab{Scarcity of Volunteers and Vantage Points.} The Open
Observatory of Network Interference (OONI)~\cite{filasto2012ooni},
ICLab~\cite{ICLab:SP20}, and Censored
Planet~\cite{Raman2020CensoredPlanet} are active censorship
measurement platforms capable of monitoring censorship across many
regions of the world. However, across all three platforms, there are
relatively few data points on Turkmenistan to provide a comprehensive
picture of the country's Web censorship. (We perform a more detailed
comparison to related work in~\sectionref{sec:relatedwork}). 
Understandably, given how slow, expensive, and strictly regulated the
Internet is in Turkmenistan, recruiting local users to run Web
connectivity tests with adequate frequency from inside the country is
difficult and potentially risky: network probes will likely stick out
from other ``allowed'' traffic. Furthermore, since VPN usage is
forbidden~\cite{hrw_TM,TM-blocks-VPNs}, a measurement system like
ICLab~\cite{ICLab:SP20} that largely depends on commercial VPNs will
have very few viable vantage points in the country for running
measurements. Further, with only 22.7K IPv4 addresses allocated for
six autonomous systems (ASes), finding enough responsive servers
(e.g., open DNS resolvers, and HTTP(S) servers) from public
infrastructure for remote measurements is unlikely to succeed.

\myparab{Measurement Machines Being Blocked.} Inspired by earlier work
on investigating the bidirectional DNS filtering in
China~\cite{GFWatch}, we tried checking which domains are censored via
DNS tampering by sending DNS queries to a non-responsive IP address
located inside Turkmenistan in late 2021. This worked successfully for
a day, but then we found that the censor stopped injecting forged
responses to our measurement machine: the IP address of our probing
machine was effectively ``banned''. Even at the time of writing this
paper, probes originated from that IP address still do not trigger any
injections. To the best of our knowledge, this is the first time we
observe an adversarial censor that intentionally hinders censorship
measurements by ignoring probing traffic for such a long time (now
more than half a year).

\myparab{Inconsistent Blocking Across Different IPs.} To reduce the
likelihood of our measurement infrastructure getting blocked, we tried
reducing the amount of probes sent to each IP address while also
originating our probes from multiple different IP addresses.
Surprisingly, even when testing from the same source IP address, we
discovered that filtering policies are not applied equally to all
destinations within Turkmenistan. Not only does censorship vary within
different IP prefixes in the same AS, but we also find variability at
the granularity of different IP addresses within the same \texttt{/24}
subnet~(discussed in \sectionref{sec:censorship-location}).
This effect is visible even in the AS (AS20661) studied in the
original Net4People discussion. If we send a censored DNS query
(\texttt{twitter.com}) to the IP address \texttt{95.85.96.36}, we can
trigger a DNS injection; however, changing the destination to the
adjacent \texttt{95.85.96.35}, we find no censorship at all.
\medskip

\noindent
Collectively, these observations point to a fascinating question: if
every IP address within Turkmenistan is potentially censored in
different ways, then how can we measure how \emph{every} IP address
with the country is being censored?
However, they also point to challenges that, when combined with those
from~\sectionref{sec:introduction}, motivated us to design a
measurement platform that can sustainably (i.e. \sysname is not
adversely affected if Turkmenistan blocks our probe machines' IP
addresses) and exhaustively measure Turkmenistan's Web censorship
infrastructure.

\definecolor{myGray}{rgb}{.7,.7,.7}
\definecolor{myBlue}{rgb}{.26,.45,.76}
\definecolor{myGreen}{rgb}{.66,.90,.65}

\begin{figure*}[t]
    \centering
    \includegraphics[width=\textwidth]{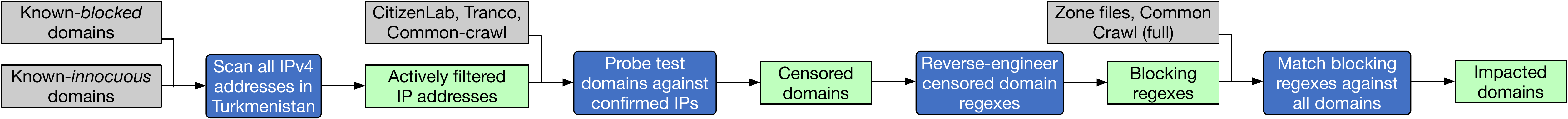}
	\caption{An overview of TMC design. {\color{myGray}Gray} boxes
	denote external datasets; {\color{myBlue}blue} boxes denote
	actions taken by \sysname; and {\color{myGreen}green} boxes denote
	\sysname's findings.}
    \label{fig:TMC}
\end{figure*}

\section{\sysname\ Design}
\label{sec:method}

Taking into account the aforementioned challenges, we design \sysname\
with the following objectives in mind. The system should be able to
\textbf{i)} confirm which IP addresses are actively being filtered,
\textbf{ii)} sustainably probe as many domains as possible across all
three protocols (DNS, HTTP, and HTTPS) to detect censored domains, and
\textbf{iii)} reverse-engineer the blocking rules of censored domains.

\subsection{Probing Mechanisms}
\label{sec:probing_mechanisms}

First and foremost, our measurement system has to achieve the above
objectives without relying on local volunteers or having access to
vantage points inside Turkmenistan. 
\sysname addresses this by sending carefully crafted probes that
elicit censorship without requiring any participation from within
Turkmenistan.

\parhead{DNS}
As shown in Figure~\ref{fig:dns_filter}, probing the DNS filter can
easily be done due to the stateless nature of DNS-over-UDP, which is
still the dominant protocol used for DNS resolutions to
date~\cite{Huston_Do53TH}.
To do so, we simply send a DNS query to an arbitrary (possibly even
unused) IP address within Turkmenistan.

\parhead{HTTP and HTTPS}
Probing the HTTP and HTTPS filters at scale is more challenging.
Recall from Figure~\ref{fig:full-handshake} that, traditionally, these
protocols require completing a TCP three-way handshake, but this would
restrict us to only studying responsive IP addresses.

In 2021, Bock et al.~\cite{Bock2021WeaponizingMF} showed that some
stateful censors could be tricked into responding without a complete
TCP handshake: by foregoing the handshake altogether and simply
sending the \texttt{PSH+ACK} containing a censored domain.
In our initial experiments, we tried to send a single \texttt{PSH+ACK}
packet with a forbidden \texttt{Host} header, but we did not get a
response from the censor.

However, after repeatedly running these tests, we discovered that by
sending one probing packet, waiting for 5 to 29 seconds, and then
sending the same probing packet again, we \emph{can} trigger both HTTP
and HTTPS censorship, as shown in
Figure~\ref{fig:incomplete-handshake}. The ``sleep'' is the time
period of the residual censorship, which is the reason why sending the
second probing packet triggers a RST from the censor.
These bounds are tight: if we sleep less than 5 seconds or more than
29, we do not observe any injected tear-down packets.

Upon further testing, we observed that the first probing packet must
have the censored domain; the second packet can be any
non-\texttt{RST}\footnote{Any TCP flag set to \texttt{PSH},
\texttt{FIN}, \texttt{URG} and/or \texttt{ACK} can trigger an
injection from both HTTP and HTTPS filters. Since our probing packets
encapsulate test domains in their application-layer payload, we opt to
use \texttt{PSH+ACK} as the flags for our measurement system so that
our traffic does not noticeably stick out from normal TCP packets that
carry data in their application-layer payload, reducing the
probability that our measurement machines will be blocked quickly.}
packet with the same TCP 4-tuple.
If the second probing packet is a \texttt{SYN+ACK} or a \texttt{SYN},
the filtering middleboxes will inject a \texttt{RST+ACK} instead of a
\texttt{RST}.

These findings align with the residual blocking behavior that we
noticed in~\sectionref{sec:Turkmenistan-censorship}.
More specifically, it appears that the first \texttt{PSH+ACK} does not
elicit a \texttt{RST}, but it does engage residual censorship, which
results in a \texttt{RST} in response to the second packet.
This corroborates one of our transport-layer evasion strategies
in~\sectionref{sec:transport-evasion}.

We believe the cause of this strange blocking behavior is the Turkmen
censor trying to be tolerant to packet loss or asymmetric routes. In
this manner, if the censor does not see the three-way handshake
complete, it will still be able to make a censorship decision if a
forbidden connection continues.
This is not unusual: prior studies have shown sophisticated censors
around the world often have more than one filtering system in place as
a backup to cope with failures of other filtering
systems~\cite{Tripletcensors, GFWatch, Bock2021EvenCH,
Bock2020DetectingAE}.

\medskip \noindent
In all the above scenarios, we could confirm that injected packets are
truly from Turkmenistan's firewalls: they all share the same
distinctive and consistent signature in the IP header with (1)~the
\texttt{IP.ID} field set to \texttt{30000} and (2)~initial
\texttt{IP.TTL} value of 128. Injected packets observed at our probing
machines will thus have \texttt{IP.TTL} values equal to 128 minus the
number of hops between our probing machines and the filtering
middlebox.

\subsection{Overall Architecture}
\label{sec:tmc_architecture}

The overall architecture of our measurement system is illustrated in
Figure~\ref{fig:TMC}, which is comprised of four main tasks.

\myparab{Confirming actively filtered IPs.} As discussed
in~\sectionref{sec:Challenges}, Turkmenistan's filtering is applied
differently at the granularity of each IP even when both filtered and
non-filtered IPs belong to the same \texttt{/24} subnet announced via
BGP. We are thus interested in examining the entire IP space of the
country to have a comprehensive view of which IPs are actively being
filtered. For this task, we obtain all IP prefixes allocated for ASes
in Turkmenistan from CAIDA's \emph{pfx2as}
dataset~\cite{CAIDA:pfx2as}. For each IP, we send packets
encapsulating opt-out, known blocked, and innocuous domains using
probing mechanisms described in~\sectionref{sec:probing_mechanisms} to
confirm which IPs are actively being filtered by Turkmenistan's
firewalls.

\parhead{Ethical Considerations.}
A primary goal of our system's design is to measure in an ethical and
responsible manner.
Unlike measurements conducted by volunteers~\cite{filasto2012ooni} or
machines that researchers can fully control~\cite{ICLab:SP20,
GFWatch}, this task involves sending probes destined for IP addresses
not under our control. 
While wide network scanning activities are common on today's
Internet~\cite{Durumeric2013ZMapFI, shodan, Bock2021WeaponizingMF},
due to the sensitive nature of censorship measurement, we follow best
practice for scanning at scale by providing an \emph{opt-out}
mechanism.
Specifically, our probes are accompanied by packets encapsulating a
non-censored domain under our control, from which our contact
information and description of the study can be found to request
opt-out. For more than two months running \sysname, we did not receive
any opt-out requests or complaints.

One may wonder whether our measurement system and evasion strategies
will help the censor to enhance its filtering capability. The general
consensus from the anti-censorship community over the years has been
that work in this space helps the evaders more than the censors. The
packet sequence used for our measurement system exploits a fundamental
aspect of the middlebox, TCP noncompliance, allowing the censor to
inject packets or block a connection even when they do not see all of
the packets in a connection~\cite{Bock2021WeaponizingMF}. This
fundamental aspect of the middlebox cannot be easily fixed.
The same reasoning applies to the evasion strategies. Censors may
patch trivial bugs, rendering a few evasion strategies ineffective.
They, however, may not be able to easily fix the fundamental problems
that enable the myriads of other strategies to succeed.

\myparab{Detecting censored domains at scale.}
Recall from \sectionref{sec:Challenges} that Turkmenistan's censorship
infrastructure ignores traffic from our measurement machines after
some
time.  To address this, we deploy our measurement machines across
different commercial virtual private servers (VPS) and frequently
change their source IP addresses.
After confirming actively filtered IP addresses in the previous task,
we distribute our probes across these confirmed IP addresses while
also scattering packets over different ports.
Designing our measurement in this fashion helps us to avoid both
(1)~false negatives due to our measurement machines being banned and
(2)~false positives due to the residual censorship applied on the same
TCP four-tuple as discussed in~\sectionref{sec:Challenges}.
We could use different port pairs for this task because Turkmenistan's
firewalls filter on all network ports, not just standard ports (i.e.,
53 for DNS, 80 for HTTP, and 443 for HTTPS).

As shown in Figure~\ref{fig:TMC}, the payload of our probes contains
domains curated from the Citizen Lab lists~\cite{CLBL}, the full
Tranco list~\cite{LePochat2019} of most popular websites, and Common
Crawl Project~\cite{CommonCrawl}. Due to limited resources of our VPS,
we opt to probe the first 10M FQDNs ranked by the Common Crawl Project
instead of the full list of almost 400M FQDNs. The rationale behind
selecting most popular domains is to shed light on the blocking status
of sites that are often visited by most Internet users. Moreover,
aggressively probing all 400M FQDNs is impractical since we do not own
the IPs being probed and would likely cause our measurement traffic to
be ignored more quickly. To that end, \textbf{we probed a total of
15.5M unique FQDNs} in September 2022.

\myparab{Reverse-engineering blocking regular expressions.} Some
initial observations reported in~\cite{TM-github-anecdotes} also
indicate that Turkmenistan's filters employ regular-expression-based
blocking. To identify these rules, once a domain is detected to be
censored by \sysname, it is broken down into substrings with different
length, prepended and appended with different random characters. These
combinations of different substrings and random characters are then
probed again to identify the shortest rule that could trigger the
filtering middlebox. For instance, when \sysname detected
\texttt{account.trendmicro.com} to be censored, our system will carry
out this task to reverse engineer the actual blocking regular
expression of
\texttt{.*\textbackslash.trendmicro\textbackslash.com.*}.

\myparab{Identifying impacted domains on the Internet.} Though we
could not probe every single domain on the Internet, it is still our
goal to assess the impact scale of Turkmenistan's
regular-expression-based censorship. After we reverse engineer the
regular expressions that the Turkmen censors use, we can test domains
offline to see if they match any of the rules. We scanned all regular
expressions that \sysname\ discovered against all FQDNs that we could
obtain from DNS zone files provided via ICANN's Centralized Zone Data
Service~\cite{ICANN} and the full host list from the Common Crawl
Project~\cite{CommonCrawl}, totaling 718M FQDNs.

\section{Who Is Being Censored?}
\label{sec:censorship-location}

We begin our analysis by investigating which IP addresses within
Turkmenistan are being subjected to censorship.

\begin{figure}[t]
\centering
\includegraphics[width=.9\columnwidth]{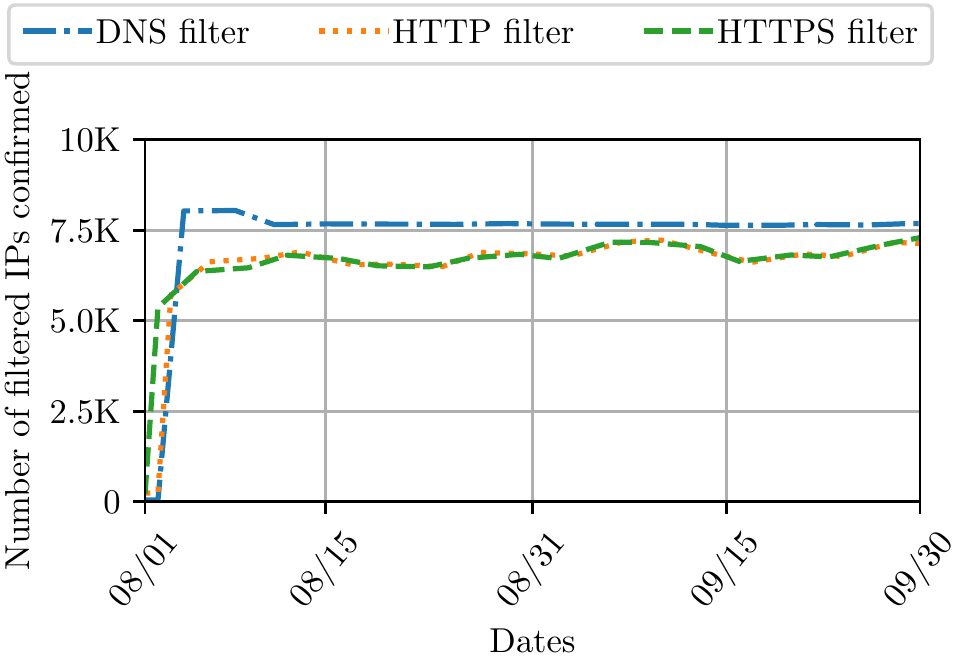}
\caption{Number of filtered IPs confirmed over time.}
\label{fig:filtered_ips}
\end{figure}

During August and September 2022, we used \sysname to scan the entire
IP address space of Turkmenistan to determine which addresses trigger
censorship for DNS, HTTP, and HTTPS. 
Figure~\ref{fig:filtered_ips} shows the total number of IP addresses
over time for each of these protocols.
The low numbers in the first few days of our measurement window are a
measurement artifact: this was before we learned Turkmenistan ignores
our measurements after a certain amount of time.
After those initial days, we switched to our distributed measurement
approach (\S\ref{sec:tmc_architecture}), which gave us consistent
results.

\begin{figure}[t]
    \centering
    \includegraphics[width=.9\columnwidth]{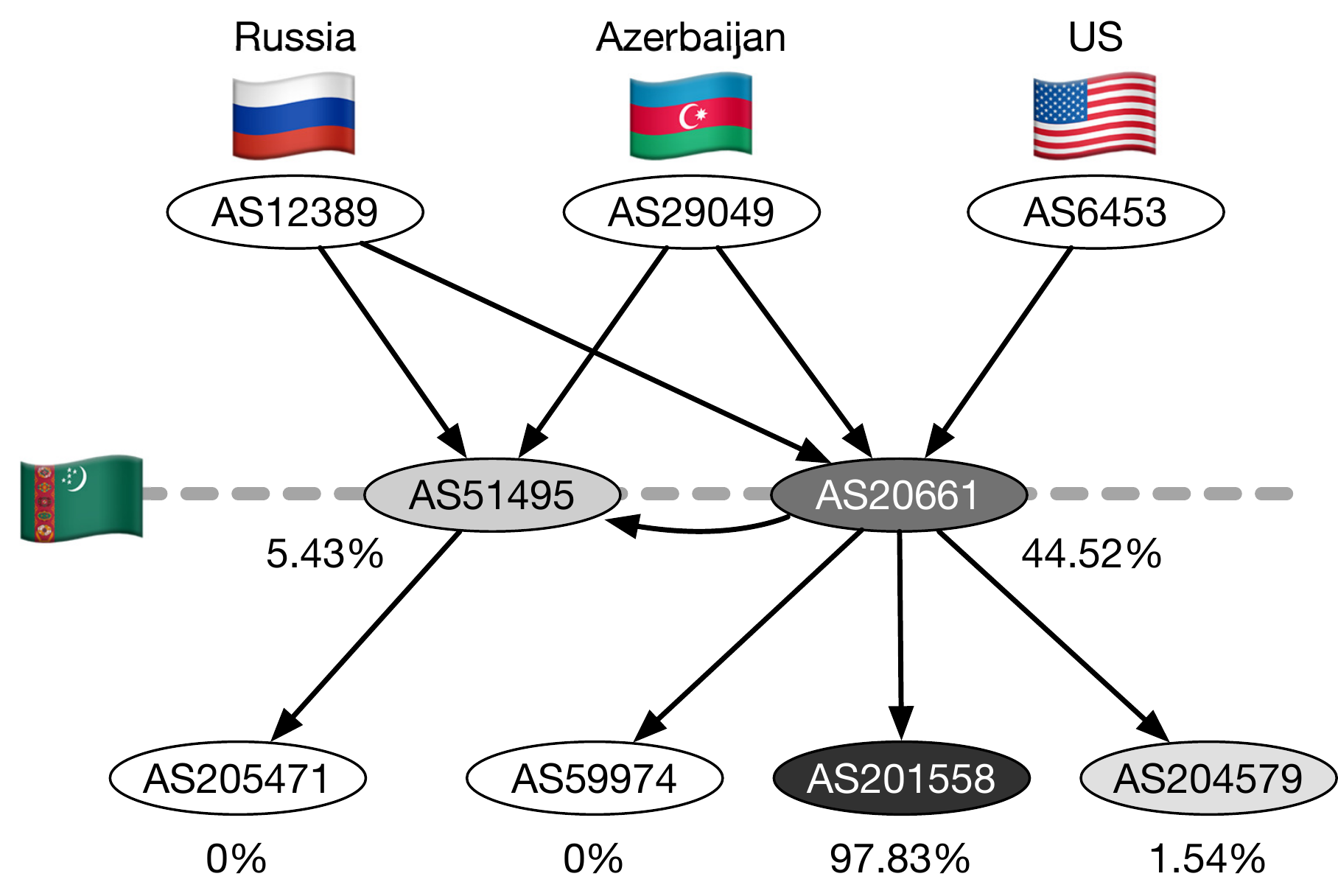}
	\caption{Turkmenistan's AS topology (edges are provider
$\rightarrow$ customer). Percentages and gray-scale denote how many of
the Turkmen AS's IP addresses are subjected to censorship.}
    \label{fig:as_topology}
\end{figure}

Figure~\ref{fig:filtered_ips} shows that \sysname could detect more
than 7.5K IPs being actively filtered on a daily basis, occupying
about 33\% of all IPs allocated for ASes in the country.
The IPs actively being filtered are similar across all three DNS,
HTTP, and HTTPS protocols.
Although the purpose of this task is to confirm which IPs are actively
filtered, we refrain from sending several probe packets to live hosts.
IPs responding to probes containing the opt-out and innocuous domains
are thus excluded from this plot and not used for later probing tasks
(\sectionref{sec:censored-domains}).
For that reason, the number of probe-able IPs confirmed via probing
HTTP and HTTPS filters are less than that of the DNS filter.

\myparab{Actively filtered IP prefixes.}
At the time of conducting our measurement, there are six ASes
allocated with a total 22.7K IPs.
These IPs are announced via 24 prefixes as shown in
Table~\ref{tab:all_prefixes} (Appendix~\ref{app:all_prefixes}).
Our probing results show that not all of these ASes are actively
filtered.
Even in ASes with IPs that \sysname\ detects to be filtered, filtering
is not applied across all addresses.

From Table~\ref{tab:all_prefixes}, we can see that the vast majority
of filtered IPs are allocated for AS20661 (State Company of Electro
Communications Turkmentelecom).
In this AS, \texttt{217.174.224.0/20} and \texttt{95.85.96.0/19} are
the two subnets with the largest number of filtered IPs (more than
6.5K IPs).
Two other ASes from which \sysname\ detects network interferences are
AS51495 (Telephone Network of Ashgabat CJSC) and AS201558 (State Bank
for Foreign Economic Affairs of Turkmenistan).

These findings explain why we initially could not trigger censorship
when probing \texttt{95.85.96.35}. This is because only 65.5\% of IPs
in \texttt{95.85.96.0/24} are actively being filtered by
Turkmenistan's firewalls. Our findings underscore the importance of
confirming actively filtered IPs to avoid false negatives when probing
against non-filtered network locations.

\myparab{AS topology.} To better understand where censorship is taking
place within Turkmenistan's network, we next look at its AS topology.
We utilize CAIDA's AS Rank~\cite{CAIDA_AS_rank} to determine
customer-provider relationships between the different ASes.
We then conduct \texttt{traceroute} for every IP prefix to obtain the
routes and routers' information via which our probing packets
traverse. To determine where network filtering happens, we use the
limited-TTL method to send multiple probe packets, encapsulating a
known censored domain, to each IP prefix. More specifically, we
incrementally increase the \texttt{IP.TTL} of our packets until we
could trigger an injection from the filtering middleboxes. Ultimately,
we were able to obtain the AS topology for Turkmenistan as shown in
Figure~\ref{fig:as_topology}.

We could identify two national gateways through which all packets
destined for any IP address in Turkmenistan will have to pass through.
One is in AS20661 and the other one is in AS51495. Our limited-TTL
experiments also indicate that the filtering middleboxes are posited
behind these gateways since our probing packets will trigger an
injection as soon as their TTL value is large enough to pass through
the gateways. We could also confirm that these filtering boxes share
the same blocking signature, i.e., the \texttt{IP.ID} field of an
injected packet is always set to \texttt{30000}
(\sectionref{sec:probing_mechanisms}).

\begin{figure}[t]
    \centering
    \includegraphics[width=.9\columnwidth]{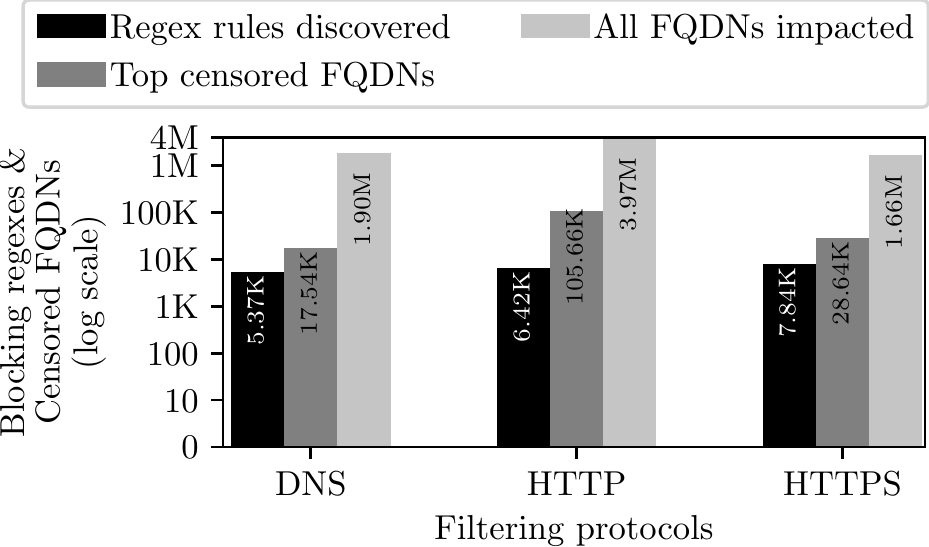}
    \caption{The number of censored FQDNs, filtering regular
     expressions, and impacted FQDNs across three protocols.}
    \label{fig:censored_domains}
\end{figure}

\section{What Is Being (Over-)Blocked?}
\label{sec:censored-domains}

Throughout the entire month of September 2022, we used \sysname to
probe (via DNS, HTTP, and HTTPS) 15.5M test domains against all of the
IP addresses we verified as being censored
(\sectionref{sec:censorship-location}).
Here, we report on what content is being blocked (and over-blocked).

\subsection{Censored Domains} 
\label{sec:block_counts}

\sysname\ detects 122K censored FQDNs, with a different number of
censored domains being detected for each protocol as shown in
Figure~\ref{fig:censored_domains}.
There are 17.54K domains filtered via DNS tampering, 105.66K domains
filtered if detected in the \texttt{Host} field of an HTTP GET packet,
and 28.64K domains filtered if detected in the SNI field of a TLS
\textit{ClientHello} packet.

\label{sec:censored_domains_classification}

\myparab{Categorization of Blocked Domains.}
To better understand the primary motivation behind Turkmenistan's
Internet censorship, we utilize Virus Total's classification
service~\cite{VirusTotal} to categorize the blocking regular
expressions that \sysname\ has discovered.
To avoid over-counting, if there are two FQDNs that match the same
inferred regular expression, we only count one of them.
For example, the blocking of \texttt{m.twitter.com} and
\texttt{www.twitter.com} should only be counted as one blocking under
the rule \texttt{.*twitter\textbackslash.com.*}.

\begin{figure}[t]
    \centering
    \includegraphics[width=\columnwidth]{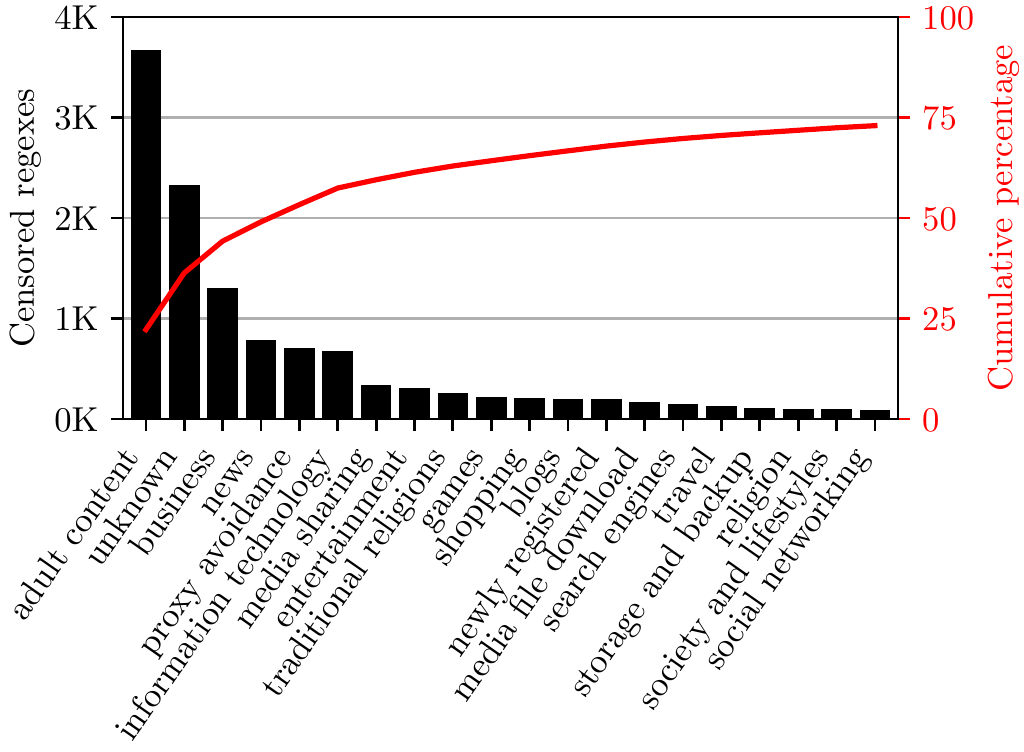}
    \caption{Top 20 domain categories of filtering regexes.}
    \label{fig:censored_categories}
\end{figure}

Figure~\ref{fig:censored_categories} shows the top categories to which
censoring regular expressions belong.
Adult Content (often pornography) alone occupies almost 25\% of all
blocking rules.
The second most dominant group of blocking rules is classified as
``unknown''.
Upon manual verifications, the vast majority of these domains are
either (1)~not hosting any Web content or (2)~currently not actively
online.
These domains might have been blocked in the past prior to our work.
Previous work has shown that once domains are added to blocklists of
nation-state firewalls they often remain blocked for an extensive
period of time regardless of their inactive status~\cite{GFWatch}.
Other top categories including typical ones that have been observed in
other countries include Business, News, and Proxy Avoidance (often
used for circumventing censorship).
Together, this group of top 20 categories makes up almost 75\% of all
blocking regular expressions.

A taxonomy of censored domains based on their language and popularity
ranking may also be desirable. A categorization based on website
language, however, is challenging as many websites serve localized
content and have different language versions for different
populations. In addition, measuring popularity is not a
straightforward task either because a meaningful popularity ranking
needs to consider several factors, including when, by who, and from
where these websites are visited.

\subsection{(Over-)Blocking Rules} 

By reverse-engineering the blocking rules of these FQDNs, we discover
16.5K unique regular expressions that define Turkmenistan's firewalls
rules.
Although the majority of these rules are ``properly'' implemented
(i.e., blocked domains are actually subdomains of the blocking rule),
there are more than about 6K rules that would cause over-blocking of
unrelated domains (i.e., blocking of domains whose second-level domain
is unrelated to the blocking regular expression).
For instance, we discover the DNS tampering rule
\texttt{.*\textbackslash.cyou.*} impacts not only the entire
\texttt{.cyou} top-level namespace with more than 770K second-level
domains but also other 117K FQDNs that happen to contain that string.

Other extreme blocking rules include
\texttt{.*wn\textbackslash.com.*}, \texttt{.*u\textbackslash.to.*},
and \texttt{.*w\textbackslash.org.*}, causing over-blocking of
hundreds of thousands unrelated domains.
The log scale of Figure~\ref{fig:censored_domains} shows the
significant magnitude of the collateral damage that these
over-blocking rules would cause, potentially affecting millions of
FQDNs.
A more detailed table listing top over-blocking rules can be found in
Table~\ref{tab:top_blocking_rules}
(Appendix~\ref{app:top_blocking_rules}).

\myparab{Unhealthy Over-blocking of DNS-over-HTTPS.}
Traditional DNS does not encrypt or authenticate its payloads, making
it trivial for censors to observe and inject lemon
responses~\cite{CollateralDamageDNS}.
To this, several domain name encryption technologies have been
proposed, e.g., DNS over TLS (DoT)~\cite{rfc7858} and DNS over HTTPS
(DoH)~\cite{rfc8484}.
By encrypting DNS traffic, these protocols take away the visibility
into plaintext DNS resolutions from the censors, effectively
circumventing their DNS filtering~\cite{Hoang2020:MADWeb,
Hoang2020:ASIACCS, Hoang2022MeasuringTA}.
Nevertheless, this has led to a new wave of blocking strategies
targeting domain name encryption protocols~\cite{Chai2019OnTI,
Basso2021MeasuringDB, Hoang2021:IP-WF, Hoang2022MeasuringTA}.
\sysname\ also detects such blocking efforts in Turkmenistan.
The DoH resolvers that \sysname\ detected also align with those
reported in the discussion of Net4People
community~\cite{TM-github-anecdotes}.

What we found intriguing, however, is the DNS filtering rule
\texttt{\^{}doh\textbackslash..*}, which we believe to be used for DoH
blocking.
Specifically, any DNS query containing a domain starting with
\texttt{\^{}doh\textbackslash.} will trigger a fake DNS injection.
While this rule is effective at blocking many publicly available DoH
resolvers~\cite{dohservers}, it also causes over-blocking of totally
unrelated websites, especially those used by Departments of Health in
many jurisdictions (e.g., \texttt{doh.gov.ae}, \texttt{doh.gov.ph},
\texttt{doh.gov.uk}, \texttt{doh.pa.gov}, \texttt{doh.wa.gov}).

\myparab{Blocking of educational domains.}
Although educational domains are not among top blocking categories,
\sysname\ found numerous domains of higher-educational institutions
being blocked, including \texttt{.*brookings\textbackslash.edu.*},
\texttt{\^{}liberty\textbackslash.edu\$}.
\texttt{\^{}hds\textbackslash.harvard\textbackslash.edu\$},
\texttt{\^{}scs\textbackslash.illinois\textbackslash.edu\$}.
Interestingly, many censored domains belong to institutions in
Michigan, such as \texttt{.*cmich\textbackslash.edu.*},
\texttt{.*wmich\textbackslash.edu.*},
\texttt{.*\textbackslash.kettering\textbackslash.edu.*}, and
\texttt{.*med\textbackslash.umich\textbackslash.edu.*}.
Based on the regular expression of each blocking rule, while some
intend to block particular departments' website many rules exhibit a
\emph{blanket} blocking behavior.


\section{Censorship Circumvention}
\label{sec:circumvention}

We discovered novel censorship evasion strategies in Turkmenistan by
using \Geneva, an open source genetic algorithm that automatically
learns by training against live censors~\cite{Bock2019GenevaEC}.
For all of our testing, we used machines under our control outside of
Turkmenistan, and issued requests to affected IP addresses inside the
country.
We started training Geneva to discover censorship evasion strategies
in March 2022. For this training, we used domains known to be censored
in Turkmenistan through anecdotal accounts \cite{TM-github-anecdotes}.
After letting \Geneva automatically discover new evasion strategies,
we performed follow-up experiments to more fully understand \emph{why}
the various evasion strategies work.

\Geneva can learn new evasion strategies by manipulating either TCP/IP
headers or application-layer payloads (specifically HTTP and DNS).
We describe these two classes of strategies in turn.

Altogether, we found five novel evasion strategies for Turkmenistan,
and re-discovered several successful strategies that prior work found
against other nation-state censors~\cite{Bock2019GenevaEC}.
All the strategies we discover work 100\% reliably.
Following precedence from prior work, we provide \Geneva's strategy
syntax; they can be copy-pasted into \Geneva's open-source engine and
used to evade censorship today.


\subsection{TCP-Based Evasion Strategies} 
\label{sec:transport-evasion}

We allowed \Geneva to train by manipulating TCP/IP headers of censored
HTTP traffic.
For every such evasion strategy \Geneva found, we also evaluated it on
HTTPS traffic.

\myparab{Strategy\ref{strat:segmentation}: TCP Segmentation}
\Geneva discovered evasion strategies that segment the TCP payload at
specific portions of the payload.
For HTTP, the strategy succeeds by segmenting the HTTP version.
For example, segmenting a \texttt{GET} request for Twitter into
(1)~\texttt{"GET / HTTP/1"}, (2)~\texttt{".1\textbackslash
r\textbackslash nUser-Agent:..."} evades censorship, but any other
segmentation index that does not split the HTTP version fails to evade
censorship.
For HTTPS, segmentation that splits the first few bytes of the TLS
Client Hello header: specifically, any segmenting at byte index 3-8
(the Record Header of the Client Hello) inclusive evades censorship.
We also find that segmentation that splits the Server Name Indication
(SNI) field evades HTTPS censorship.

\begin{strategy}
\begin{verbatim}
 [TCP:flags:PA]-fragment{tcp:8:True}-| \/
\end{verbatim}
	\caption{\bf (HTTP) Segmentation (Similar for HTTPS)}
\label{strat:segmentation}
\end{strategy}

We believe this strategy works by interfering with the censor's
internal parsing used to identify the request as HTTP or HTTPS. Since
the censor fails open, the request is allowed to pass through.

\myparab{Strategy~\ref{strat:tcb-teardown}: TCB Teardown}
One of the most famous packet manipulation layer strategies is the TCB
Teardown~\cite{Khattek2013,ZWang2017INTANG,ZWang2020SymTCP,Bock2019GenevaEC}.
A client performs this strategy by injecting a \texttt{RST} or
\texttt{FIN} packet in such a way that it is not processed by the
destination server, often by limiting the \texttt{TTL} (time-to-live)
field.
The censor processes the teardown packet, incorrectly assumes that the
connection has been torn down, and stops tracking the connection,
enabling the client and server to continue communicating censorship
free.

\begin{strategy}
\begin{verbatim}
 [TCP:flags:S]-duplicate(,
   duplicate(tamper{TCP:flags:replace:R}(
     tamper{TCP:chksum:corrupt},),))-| \/
\end{verbatim}
	\caption{\bf (HTTP \& HTTPS) TCB Teardown via RST}
\label{strat:tcb-teardown}
\end{strategy}

\Geneva discovered such strategies in Turkmenistan.
Strategy~\ref{strat:tcb-teardown} sends a \texttt{RST} packet with a
broken checksum immediately after sending the \texttt{SYN} packet.
\Geneva identified other variants of this strategy, including tearing
down with a \texttt{FIN} packet. 
Any combination of flags that includes a \texttt{RST} or a
\texttt{FIN} is sufficient to evade censorship, including nonsensical
flag combinations like \texttt{PSH+RST+FIN+SYN}.
These evade censorship for HTTP and HTTPS.

We also tested the TCB Teardown strategies discovered by Bock et
al.~\cite{Bock2019GenevaEC}, and found that all of them are successful
against Turkmenistan's HTTP and HTTPS censorship.

\myparab{Strategy~\ref{strat:free-pass}: Free Pass}
\Geneva discovered a novel strategy against Turkmenistan's censor.
This strategy sends a \texttt{RST} or \texttt{FIN} packet
\emph{before} sending the initial \texttt{SYN}.
At first, this strategy appears to be a TCB Teardown attack, but it is
not: since the \texttt{SYN} packet comes after the \texttt{RST}, the
censor should not yet have any state to tear down. 
Further, we find that the injected \texttt{RST} or \texttt{FIN}
packets only stave off censorship if the \texttt{SYN} packet is sent
less than 5 seconds later.
If the \texttt{SYN} is sent on or after the fifth second, the strategy
no longer works. 
This strategy works for both HTTP and HTTPS. 

\begin{strategy}
\begin{verbatim}
 [TCP:flags:S]-duplicate(
   tamper{TCP:flags:replace:R},)-| \/
\end{verbatim}
	\caption{\bf (HTTP \& HTTPS) Free Pass (for < 5 seconds)}
\label{strat:free-pass}
\end{strategy}

Frankly, we do not understand why this strategy works. 
We initially hypothesized that the \texttt{RST} and \texttt{SYN}
packets might be arriving at the censor in the wrong order, causing a
normal TCB teardown attack. 
But this is not the case: we can delay the \texttt{SYN} packet by up
to 5 seconds after the \texttt{RST} packet, and we can still avoid
censorship.
These timing dynamics are also not present in the normal TCB teardown
strategy. 
In the normal TCB Teardown case, the \texttt{RST} and \texttt{FIN}
packets are effective for more than 5 seconds; we tested this by
delaying the forbidden request after injecting the teardown packet.

The timing dynamics of this strategy mirror the dynamics we discovered
with our measurement strategy: we do not receive censored responses
from the server when we send an incomplete TCP handshake for 5 seconds
from the first incomplete handshake, but get censored on the 5th
second.

Amazingly, the client does not have to send a \texttt{RST} or
\texttt{FIN} packet in order to evade censorship with this strategy:
the client may simply elicit a \texttt{RST} from the server. 
This can be done by sending an innocuous \texttt{PSH+ACK} to the
server \emph{before} a TCP handshake has been established, causing the
server to respond with a \texttt{RST}.
This strategy suggests that there may be simple server-side evasion
strategies~\cite{Bock2020ServerSide} that are successful against
Turkmenistan's censors.

\subsection{Application-Layer Evasion Strategies} 
\label{sec:application-evasion}

We also trained \Geneva by manipulating DNS and HTTP payloads.

\myparab{DNS Strategies: Elevated Count Above a Threshold}
When we set the \texttt{qdcount}, \texttt{ancount}, \texttt{nscount},
and/or \texttt{arcount} fields within the DNS query header to values
above a certain threshold, we are able to bypass Turkmenistan's DNS
injection. Through empirical testing, we determined that the threshold
for all of these fields is 25. Although this elevated count is in
violation of the RFC, many DNS servers still respond to such queries
(as discovered previously~\cite{Harrity2022}).

\begin{strategy}
\begin{verbatim}
 [DNS:*:*]-tamper{DNS:ancount:replace:32}-| \/
\end{verbatim}
\caption{\bf (DNS) Elevated Count Above a Threshold}
\label{strat:dns-count}
\end{strategy}

We note that the censor does not inspect DNS response packets, so if
the DNS request itself is not censored, the request should succeed.

We also found that the DNS censor changed during the course of our
study.
\Geneva initially found a simple strategy that created a second DNS
question record without incrementing the query count
(\texttt{qdcount}) field. This strategy evaded censorship: we would
not get an injection with an A record pointing to 127.0.0.1, but
instead a corrupted injection with an A record pointing to 0.0.1.44
(which a normal client would ignore). After further analysis, we
realized that the response the censor had sent was malformed. The
injected response had two answer records, but the response's
\texttt{arcount} field was set to 1. This gave way to a parsing error,
which made it seem as if the DNS injection's A record was pointing to
another IP address, when in reality the injection had two A records,
both pointing to 127.0.0.1.

Interestingly, Turkmenistan caught their mistake during the course of
our study. Around May 26th, 2022, Turkmenistan fixed the censor so
that their responses would have the correct number of answer records.
Now, the censor responds to \Geneva's request with one A record
pointing to 127.0.0.1 because the request declared a \texttt{qdcount}
of 1.  As a result, this strategy no longer works. 

\myparab{Strategy~\ref{strat:http-whitespace}: HTTP Host Header
Whitespace}
\Geneva discovered one novel HTTP strategy that bypasses censorship by
inserting additional whitespace within the host header.  More
specifically, the strategy inserts a tab and a new line right before
the host header value.

\begin{strategy}
\begin{verbatim}
 [HTTP:host:*]-insert{%09%0A:start:value:1}-| \/
\end{verbatim}
\caption{\bf (HTTP) Host Header Whitespace}
\label{strat:http-whitespace}
\end{strategy}

Harrity et al.~\cite{Harrity2022} discovered 77 application-layer HTTP
strategies that evade censorship in other nation-states, and we tested
each against Turkmenistan. We find only 5 of these were successful;
see the detailed breakdown in the appendix.


\section{Related Work}
\label{sec:relatedwork}

\Parhead{Measurement of Turkmenistan's Censorship.}
Over the past decade, there have been several efforts to measure
censorship in Turkmenistan.
Some of these focused specifically on Turkmenistan~\cite{OSF_study,
Qurium, ValdikSS} while others performed global-scale
measurements~\cite{filasto2012ooni, Satellite, ICLab:SP20}.
Table~\ref{tab:comparison_table} presents a detailed comparison
between our study and these previous efforts, showing how \sysname's
unique measurement method has allowed us to gain a more comprehensive
view into Turkmenistan's Internet censorship landscape only after a
short period of time.

\begin{table}[t]
    \centering
    \caption{Comparison between different Turkmenistan censorship measurement
    studies/platforms.}
    \resizebox{\columnwidth}{!}{
    \begin{tabular}{llcrrrr}
    \toprule
    \textbf{Study}                         &\textbf{When}    & \textbf{Censored/Tested}  & \textbf{Method}        & \textbf{Coverage} \\ \hline
    SecDev~\cite{OSF_study}                 & 07/12--08/12   & 34/unknown        & local                   & unknown           \\
    Qurium~\cite{Qurium}                    & 07/19             & 133/10K           & unknown                 & unknown           \\
    ValdikSS~\cite{ValdikSS}                & 12/18/21        & 60/1K             & unknown                 & 1 AS              \\
    OONI (DNS)~\cite{filasto2012ooni}       & 02/17--09/22   & 254/2.2K          & volunteers              & 5 ASes            \\
    Satellite~\cite{Satellite}              & 08/18--05/22   & 267/4.7K          & open resolvers          & 2 ASes            \\
    \sysname                                & 09/22             & 122K/15.5M        & \emph{No} vantage point & All ASes          \\
    \bottomrule
    \end{tabular}}
    \label{tab:comparison_table}
  \end{table}

\parhead{Remotely Measuring Censorship.}
There have been myriad prior efforts to develop techniques that allow
one to measure censorship of a country without requiring a vantage
point from inside that country.
The CensoredPlanet platform~\cite{Raman2020CensoredPlanet}
incorporates multiple techniques, such as Quack~\cite{quack}, that can
remotely measure Internet censorship without participating users in
the country.
However, these generally require some form of responsive server
(typically echo servers) within the country.
Such servers are unfortunately not widely available in countries with
low Internet penetration like Turkmenistan.
\sysname employs a novel sequence of packets that trigger censorship
without requiring any server-side participation within the country;
while this borrows techniques from Bock et
al.~\cite{Bock2021WeaponizingMF}, we believe we are the first to apply
them to wide-scale detection of censorship.

\parhead{Evading Turkmenistan's Censorship.}
While there has been some earlier effort in the community to manually
craft packets to sidestep Turkmenistan
censorship~\cite{TM-github-anecdotes}, we are not aware of any prior
studies that have systematically investigated censorship circumvention
across different network layers in Turkmenistan.
While other general mechanisms work to varying degrees of success,
such as tunneling censored traffic over anonymity networks (e.g.,
Tor~\cite{Dingledine:Usec13},
I2P~\cite{Hoang2018:i2p-censorship-resistance}), we believe we are the
first to find circumvention strategies specific to their censorship
infrastructure.

\section{Conclusion}
\label{sec:conclusion}

In this paper, we presented the most thorough evaluation of
Turkmenistan's censorship of the Web (DNS, HTTP, and HTTPS).
We found that blocking is not applied to all IP addresses equally, and
that there are millions of domains that are very likely over-blocked
due to inaccurate regular-expression rules.
These results would not have been possible with traditional
measurement techniques, which require some degree of user
participation or server availability within the censored country.
The design of \sysname has enabled us to perform a large-scale
measurement in a low-Internet-penetration country like Turkmenistan.
While the specific packet sequences designed for \sysname may not work
elsewhere, the high-level approach can. Our study is a first step
towards country-wide measurements from the outside without access to
responsive vantage points or volunteers. We hope that our paper will
lead to more work in this direction.
To assist in such future efforts, we make our measurement code and the
evasion strategies discovered by Geneva to aid in its integration with
any existing evasion software, publicly available at
https://doi.org/10.5281/zenodo.7631411.

\section*{Acknowledgements}

We would like to thank all the anonymous reviewers for their thorough
feedback. We also thank Sudhamshu Hosamane, Martin Lutta Putra, and
others who preferred to remain anonymous for helpful comments and
suggestions to improve this paper.

This research was supported in part by  the National Science
Foundation (NSF) through award CNS-1943240, and the Defense Advanced
Research Projects Agency (DARPA) through award HR00112190126. The
opinions in this paper are those of the authors and do not necessarily
reflect the opinions of the sponsors.


\bibliographystyle{ACM-Reference-Format}
\balance
\bibliography{conferences,refs}
\appendix
\section{Turkmenistan's Actively Filtered IP Addresses}
\label{app:all_prefixes}

Table~\ref{tab:all_prefixes} lists all IPv4 prefixes observed via BGP
announcements, their organization information, and the average
percentage of actively filtered IP addresses that \sysname\ has
detected over time. Note that there are some \texttt{/24} prefixes
belong to less specific ones (e.g., \texttt{/20}, \texttt{/22}). The
majority of actively filtered IPs are within the state-controlled
AS20661.

\begin{table}[t]
      \caption{IPv4 prefixes allocated to Turkmenistan ASes and the
      average percentage of filtered IPs observed over time.}
      \label{tab:all_prefixes}
      \centering

      \resizebox{\columnwidth}{!}{%
      \begin{tabular}{|l|l|l|c|}
      \hline
      \multicolumn{1}{|c|}{A}      & {IP prefix}        & {Filtered (\%)}   & {Organization} \\ \hline
      \multirow{15}{*}{AS20661}    & 103.220.0.0/22     & 0                 & \multirow{7}{*}{\thead{State Company of\\Electro Communications\\Turkmentelecom}} \\
                                   & 119.235.112.0/20   & 0.34              & \\
                                   & 177.93.143.0/24    & 0.41              & \\
                                   & 185.69.184.0/24    & 0                 & \\
                                   & 216.250.8.0/21     & 0                 & \\
                                   & 217.174.224.0/20   & \textbf{54.02}    & \\
                                   & 95.85.96.0/19      & \textbf{53.47}    & \\\cline{2-4}
                                   & 95.85.100.0/22     & 9.58              & \multirow{8}{*}{\thead{Leased line customers}}   \\
                                   & 95.85.100.0/24     & 9.45              & \\
                                   & 95.85.101.0/24     & 9.52              & \\
                                   & 95.85.104.0/22     & 11.13             & \\
                                   & 95.85.104.0/24     & 11.33             & \\
                                   & 95.85.96.0/24      & \textbf{65.55}    & \\
                                   & 95.85.98.0/24      & 0.065             & \\
                                   & 95.85.99.0/24      & 0                 & \\ \hline
      \multirow{5}{*}{AS51495}     & 93.171.220.0/22    & 5.43              & \multirow{5}{*}{\thead{Telephone Network of\\Ashgabat CJSC}}  \\
                                   & 93.171.220.0/24    & 0                 & \\
                                   & 93.171.221.0/24    & 0                 & \\
                                   & 93.171.222.0/24    & 0                 & \\
                                   & 93.171.223.0/24    & \textbf{21.72}    & \\ \hline
      AS205471                     & 185.69.185.0/24    & 0                 & \thead{State Company of\\Electro Communications\\Turkmentelecom}  \\ \hline
      AS59974                      & 185.69.186.0/24    & 0                 & \thead{Mobile Customers\\Inet Access}  \\ \hline
      AS201558                     & 185.69.187.0/24    & \textbf{97.83}    & \thead{State Bank for\\Foreign Economic Affairs\\of Turkmenistan}  \\ \hline
      AS204579                     & 185.246.72.0/22    & 1.54              & \thead{Turkmen hemrasi CJSC}  \\ \hline
      \end{tabular}
        }
    \end{table}

\section{Top Blocking Rules and Sample Censored FQDNs}
\label{app:top_blocking_rules}

Table~\ref{tab:top_blocking_rules} shows the top ten overblocking
rules with some sample impacted FQDNs. We can see that some blocking
regular expressions such as \texttt{.*\textbackslash.cyou.*},
\texttt{.*porn.*}, and \texttt{.*\textbackslash.rocks.*} are all
active generic top-level domains (gTLDs). These blocking rules, thus,
would block access to not only all second-level domains (SLDs)
registered under these gTLDs but also other SLDs happen to contain
such strings. In addition, extremely short blocking rules like
\texttt{.*w\textbackslash.org.*} (a WordPress domain) tends to cause
large collateral damage as they end up blocking totally unrelated yet
valuable domains (e.g., \texttt{tensorflow.org}). Even
internationalized domains, e.g., xn--vl2b99byzlcpd9yy.com, which is a
Korean website unrelated to \texttt{.*yy\textbackslash.com.*}, are
impacted by these short blocking rules.

\begin{table}[t]
  \caption{Top 10 blocking regular expressions with the highest number of impacted FQDNs.}
  \label{tab:top_blocking_rules}
  \centering

  \resizebox{\columnwidth}{!}{%
  \begin{tabular}{|r|c|r|}
  \hline
  \small{\thead{\# impacted\\FQDNs}}           & \small{\thead{Regular expressions\\(Blocking protocols)}}   & \small{Sample domains} \\
  \hline
  887K  & \texttt{.*\textbackslash.cyou.*}     & \texttt{cyou-TLD zone}, \texttt{movizland.cyou}, \texttt{starlink.cyou}      \\
        & \texttt{DNS, HTTPS}            & \texttt{committee.cyou}, \texttt{gomovies.cyou}     \\
  \hline
  568K  & \texttt{.*vpn.*}                     & \texttt{nordvpn.com}, \texttt{avira-vpn.com}, \texttt{expressvpn.com}\\
        & \texttt{HTTP}                 & \texttt{openvpn.net}, \texttt{vpnoverview.com}, \texttt{vpnmentor.com}     \\
  \hline
  480K  & \texttt{.*porn.*}                    & \texttt {porn-TLD zone}, \texttt{pornhub.com}, \texttt{youporn.com},       \\
        & \texttt{DNS}                         & \texttt{nopornnorthampton.org}, \texttt{pornphiphit.co.th}     \\
        &                                      & \texttt{antipornography.org}     \\
  \hline
  300K  & \texttt{.*wn\textbackslash.com.*}    & \texttt{wn.com}, \texttt{423down.com}, \texttt{dawn.com}    \\
        & \texttt{HTTP}                        &  \texttt{uptodown.com}, \texttt{respawn.com}, \texttt{bandsintown.com}     \\
  \hline
  267K  & \texttt{.*u\textbackslash.to.*}      & \texttt{u.to}, \texttt{shahed4u.town}, \texttt{u.today}      \\
        & \texttt{HTTP}                        & \texttt{hindilinks4u.to}, \texttt{dsu.toscana.it}    \\
  \hline
  217K  & \texttt{.*w\textbackslash.org.*}     & \texttt{cookielaw.org}, \texttt{w.org}, \texttt{hrw.org}       \\
        & \texttt{HTTP}                        & \texttt{jw.org}, \texttt{democracynow.org}, \texttt{tensorflow.org}      \\
  \hline
  208K  & \texttt{.*xx\textbackslash.com.*}    & \texttt{code-boxx.com}, \texttt{uproxx.com}, \texttt{mixx.com}      \\
        & \texttt{HTTP}                        & \texttt{idexx.com}, \texttt{tkmaxx.com}, \texttt{speexx.com}     \\
  \hline
  192K  & \texttt{.*\textbackslash.rocks.*}    & {rocks-TLD zone}, \texttt{vox.rocks}, \texttt{say.rocks}      \\
        & \texttt{DNS, HTTPS}                  & \texttt{kavin.rocks}, \texttt{yolk.rocks}     \\
  \hline
  175K  &\texttt{.*twitter\textbackslash.com.*}& \texttt{bretwitter.com}, \texttt{notrealtwitter.com}      \\
        & \texttt{DNS, HTTP, HTTPS}            & \texttt{spotwitter.com}, \texttt{johnnys-twitter.com}     \\
        &                                      & \texttt{financetwitter.com}     \\
  \hline
  173K  & \texttt{.*yy\textbackslash.com.*}    & \texttt{ammyy.com}, \texttt{abbyy.com}, \texttt{haliyy.com}      \\
        & \texttt{HTTP}                        & \texttt{playdayy.com}, \texttt{xn--vl2b99byzlcpd9yy.com}     \\
  \hline
  \end{tabular}
    }
\end{table}

\section{Application-Layer HTTP Strategies}
Harrity et al.~\cite{Harrity2022} recently discovered 77
application-layer HTTP using open-source modifications to \Geneva.
We tested all of those strategies, and report the successful 5
strategies here.
Coincidentally, all five of these strategies were discovered against
Kazakhstan's HTTP censorship.
Note that although the same strategies work here as in Kazakhstan,
they may not succeed for the same reason.

\bigbreak
\noindent
\begin{tabular}{p{.96\columnwidth}}
    \specialrule{.075em}{.05em}{.05em}
    \textbf{HTTP Strategy 1: Request Line Whitespace After Version}\\
    \specialrule{.075em}{.05em}{.05em}
    \texttt{[HTTP:version:*]-insert\{\%20\%0A\%09:end:value:1\}-| \textbackslash/} \\
    \specialrule{.075em}{.05em}{.05em}
\end{tabular}
\bigbreak

This strategy inserts a space, newline, and tab right after the HTTP
version. Other spaces, newlines, and/or tabs may be inserted after the
version as well, as long as these three characters are in the correct
sequence. Any omission or swapping of the three characters causes the
strategy to fail. However, a variant of this strategy, where one or
more spaces are inserted after the version, succeeds. 

\bigbreak
\noindent
\begin{tabular}{p{.96\columnwidth}}
    \specialrule{.075em}{.05em}{.05em}
    \textbf{HTTP Strategy 2: Request Line Whitespace After Method}\\
    \specialrule{.075em}{.05em}{.05em}
    \texttt{[HTTP:method:*]-insert\{\%0A:start:value:1\}-| \textbackslash/} \\
    \specialrule{.075em}{.05em}{.05em}
\end{tabular}
\bigbreak

This strategy inserts a new line right before the HTTP method. The
strategy succeeds with any number of new lines inserted before the
method as long as there is at least one. 

We believe these strategies work by breaking how the Turkmenistan
censor identifies the HTTP version, which we know the censor relies on
to parse.

\bigbreak
\noindent
\begin{tabular}{p{.96\columnwidth}}
    \specialrule{.075em}{.05em}{.05em}
    \textbf{HTTP Strategy 3: Sandwich Strategy Ver. 1}\\
    \specialrule{.075em}{.05em}{.05em}
    \texttt{[HTTP:host:*]-insert\{\%20:end:value:3391\}(\newline
            \hspace*{3em}duplicate(duplicate(,\newline 
            \hspace*{6em}replace\{a:name:1\}),\newline 
            \hspace*{9em}insert\{\%09:start:name:1\}),)-| \textbackslash/} \\
    \specialrule{.075em}{.05em}{.05em}
\end{tabular}
\bigbreak

The most complex strategy discovered by Harrity et
al.~\cite{Harrity2022} is the \emph{sandwich strategy}. This strategy
works by inserting specially crafted HTTP headers before and after the
forbidden HTTP header to prevent the censor from reading the HTTP host
successfully.
This specific strategy has two components; first, it inserts 3,391 (or
more) spaces after the host header value. Second, it creates new HTTP
headers before and after the Host header (but with a tab character
before the trailing header).
We modified this strategy from Harrity's original; through manual
experimentation, we discovered that 3,391 spaces is the minimum
successful number to evade the Turkmenistan censor. 
If less than 3,391 spaces are inserted, the strategy does not evade
censorship. 
We initially hypothesized this strategy was inducing TCP segmentation
by increasing the packet size; but we find that when TCP segmentation
occurs, it does not occur at indices that evade censorship when the
spaces are omitted. 
Instead, we hypothesize this strategy is working by exceeding the
maximum number of bytes that the Turkmenistan censor can process.

Here is a second variant of the strategy, which works the same way.

\bigbreak
\noindent
\begin{tabular}{p{.96\columnwidth}}
    \specialrule{.075em}{.05em}{.05em}
    \textbf{HTTP Strategy 4: Sandwich Strategy Ver. 2}\\
    \specialrule{.075em}{.05em}{.05em}
    \texttt{[HTTP:host:*]-insert\{\%20:end:value:3391\}(\newline
            \hspace*{3em}duplicate(duplicate(\newline
            \hspace*{6em}insert\{\%09:start:name:1\},),\newline
            \hspace*{9em}replace\{a:name:1\}),)-|\textbackslash/} \\
    \specialrule{.075em}{.05em}{.05em}
\end{tabular}
\bigbreak
\end{sloppypar}

\end{document}